\documentclass[fleqn,reqno]{amsart}

\usepackage{graphicx,amscd,amsmath,amssymb,verbatim}
\usepackage[dvips]{hyperref}
\usepackage[TS1,OT1,T1]{fontenc}

\begin{document}

\title[Hamilton relativity in quantum mechanics]{Hamilton relativity group for noninertial states in quantum mechanics}
\author{Stephen G. Low}
\address{Austin, Texas }
\email{Stephen.Low@alumni.utexas.net}
\date{\today}
\subjclass[2000]{}
\maketitle
\begin{abstract}
Physical states in quantum mechanics are rays in a Hilbert space.
Projective representations of a relativity group transform between
the quantum physical states that are in the admissible class. The
physical observables of position, time, energy and momentum are
the Hermitian representation of the generators of the algebra of
the Weyl-Heisenberg group.\ \ We show that there is a consistency
condition that requires the relativity group to be a subgroup of
the group of automorphisms of the Weyl-Heisenberg algebra.\ \ This,
together with the requirement of the invariance of classical time,
results in the inhomogeneous Hamilton group. The Hamilton group
is the relativity group for noninertial frames in classical Hamilton's
mechanics. The projective representation of a group is equivalent
to unitary representations of the central extension of the group.
The central extension of the inhomogeneous Hamilton group and its
corresponding Casimir invariants are computed. One of the Casimir
invariants is a generalized spin that is invariant for noninertial
states.\ \ It is the familiar inertial Galilean spin with additional
terms\ \ that may be compared to noninertial experimental results.
\end{abstract}
\section{Introduction}

A relativity group defines a universal transformation between physical
states in an admissible class. For special relativity, the inhomogeneous
Lorentz group transforms\ \ inertial states that have relative rotation,
velocity and translations in position and time\ \ into one another.
It specifies how the measuring rods, clocks, momentum meters and
energy meters of these states are related.\ \ For small velocities,
the Lorentz group is approximated by the Euclidean group of Galilean
relativity.\ \ In the quantum formulation, the physical states are
realized as rays in a Hilbert space and the group acts through a
projective representation. This is equivalent to the unitary representation
of the central extension, both algebraic and topological, of the
relativity group. The Poincar\'e group, that is the central extension
of the inhomogeneous Lorentz group, is simply its cover as it does
not have an algebraic extension \cite{Weinberg1}.\ \ The Galilei
group is the central extension of the inhomogeneous Euclidean group
that has a central mass generator as an algebraic extension. 

We review projective representations on physical states that are
rays in a Hilbert space and then show that this property of quantum
mechanics leads directly to a\ \ consistency condition that relativity
groups must be subgroups of the group of automorphisms of the Weyl-Heisenberg
algebra. The Hamilton group is the subgroup of this group of automorphisms
that leaves time invariant.\ \ In a previous paper it was shown
that the Hamilton group is the relativity group for noninertial
frames in classical Hamilton's mechanics \cite{Low7}.\ \ The projective
representations of the inhomogeneous Hamilton group must be determined
for its action on a quantum noninertial state. We will show that
the central extension of the Hamilton group admits three central
generators; $I$ for the position-momentum and time-energy quantum
commutation relations, $M$ as the mass generator and a new central
element $A$ with dimensions that are the reciprocal of tension.
The Galilei group is the subgroup of this group for inertial states
and the mass generator is the same. The Casimir invariants are calculated
and this leads to a generalized definition of Galilei spin invariant
for noninertial physical states. 
\section{Projective representations of groups in quantum mechanics}

Physical states are represented in quantum mechanics by rays $\Psi
$ in a Hilbert space $\text{\boldmath $\mathrm{H}$}$\ \ \ Rays are
equivalence classes of states $|\psi \rangle \in \text{\boldmath
$\mathrm{H}$}$ that differ only in phase \cite{bargmann},\cite{mackey2}.\ \ Two
states $|\psi \rangle ,|\tilde{\psi }\rangle $ are in the same equivalence
class $\Psi $ if 
\begin{equation}
\left. \tilde{\left. \left| \psi \right. \right\rangle  } = e^{i
\omega } \left| \psi \right. \right\rangle  , \omega \in \mathbb{R}\text{}.
\end{equation}

A symmetry of the physical system, described by a Lie group $\mathcal{G}$
with elements $g$, acts on the rays through a projective representation
$\pi $ to transform one physical state into another. 
\begin{equation}
 \tilde{\Psi } = \pi ( g)  \Psi .
\end{equation}

The projective representation has the property\ \ $\pi ( g) \pi
( \tilde{g}) =e^{i \omega ( g,\tilde{g}) }\pi ( g \tilde{g}) $,
$\omega ( g,\tilde{g}) \in \mathbb{R}$. As the rays are equivalence
classes, $\Psi \simeq e^{i \omega ( g,\tilde{g}) }\Psi $\ \ for
any $\omega ( g,\tilde{g}) \in \mathbb{R}$ and therefore
\begin{equation}
\pi ( g) \pi ( \tilde{g}) \Psi  =\pi ( g \tilde{g})  \Psi \text{}.
\end{equation}

A class of observables is defined by lifting the projective representation
to act on elements $X$ of the Lie algebra $\text{\boldmath $\mathrm{a}$}(
\mathcal{G}) $ so that $\pi ^{\prime }( X) =(T_{e}\pi )(X)$ is an
operator on $\text{\boldmath $\mathrm{H}$}$.\ \ \ An observable
$\hat{\tilde{X}}$ acting on the state $\tilde{\Psi } $ is related
to the observable $\hat{X}$ acting on the state $\Psi $ by the projective
representation of $g$\ \ 
\begin{equation}
\pi ( g)  \hat{X} \Psi =\pi ( g)  \hat{X} {\pi ( g) }^{-1}\pi (
g)  \Psi =\text{}\pi ( g)  \hat{X} {\pi ( g) }^{-1} \tilde{\Psi
} =\pi ^{\prime }( \tilde{X}) \tilde{\Psi }=\hat{\tilde{X}} \tilde{\Psi
} .%
\label{transform state}
\end{equation}

\noindent Therefore 
\begin{equation}
\pi ^{\prime }( \tilde{X}) =\pi ( g)  \pi ^{\prime }( X)  {\pi (
g) }^{-1}=\pi ^{\prime }( g X g^{-1}) 
\end{equation}

\noindent for which a sufficient condition is\footnote{If the representation
is an isomorphism, then it is also necessary.\ \ Otherwise, equivalence
is defined up to the equivalence class defined by taking the quotient
of the group with the kernel of the representation.}
\begin{equation}
\tilde{X}=g X g^{-1}.%
\label{transform generator}
\end{equation}

\noindent Finally, we note that if an operator $X$ is an invariant
such that $\hat{\tilde{X}} =\hat{X}$, then 
\begin{equation}
X=g X g^{-1}.
\end{equation}

A theorem in representation theory states that the projective representations
$\pi $ of a Lie group $\mathcal{G}$ are equivalent to the unitary
or anti-unitary\footnote{In what follows we will use only unitary
with anti-unitary implicitly included} representation $\varrho $
of the central extension $\check{\mathcal{G}}$.\footnote{If $\mathcal{G}\simeq
\check{\mathcal{G}}$, the group does not have any intrinsic projective
representations and the projective representations are just the
unitary representations. } (See Section 2.7 and Appendix B of \cite{Weinberg1}.)\ \ Therefore,
the physical system may be studied by characterizing the unitary
irreducible representations $\varrho$ of\ \ the group $\check{\mathcal{G}}$
acting on a Hilbert space ${\text{\boldmath $\mathrm{H}$}}^{\varrho
}$. The Hilbert space is labeled by the unitary irreducible representation
because it is not given {\itshape a priori}, but rather is determined
by the unitary representation.\ \ 

The unitary representations act on the states $|\psi \rangle $ in
the Hilbert space ${\text{\boldmath $\mathrm{H}$}}^{\varrho }$ 
\begin{equation}
\tilde{\left. \left| \psi \right. \right\rangle  } =\ \ \varrho
( g)  \left| \psi \right\rangle  ,\ \ g\in \check{\mathcal{G}}.
\end{equation}

\noindent The unitary representation $\varrho $ may be lifted to
the tangent space to define the Hermitian representation $\hat{X}$
of the element of the algebra $X$ 
\begin{equation}
\hat{X}=\varrho ^{\prime }( X) =T_{e}\varrho ^{\prime }( X) .
\end{equation}

Physical observables are characterized by the eigenvalues of the
Hermitian representation of the generators
\begin{equation}
\left. \left. \hat{X}\left| \psi \right. \right\rangle  =x\left|
\psi \right. \right\rangle  ,\ \ \ x\in \mathbb{R}\text{},
\end{equation}

\noindent and these generators transform as 
\begin{equation}
\varrho ( g) \hat{X}\left| \psi \right\rangle  =\varrho ( g) \hat{X}{\varrho
( g) }^{-1} \tilde{\left.  \left| \psi \right. \right\rangle  }=\hat{\tilde{X}}\tilde{\left.
\left| \psi \right. \right\rangle  }.%
\label{generator transform under group}
\end{equation}
\section{Quantum mechanics consistency condition with a relativity
group}

Measurements of the basic physical observables such as position,
time, energy and momentum depend on the relative physical state
in which they are measured. For certain classes of physical states,
there is a relativity principal that defines a universal group relating
the states. Examples were given in the introduction: the inhomogeneous
Lorentz group, and its central extension the Poincar\'e group for
the class of inertial quantum states in special relativistic quantum
mechanics \cite{Weinberg1} and the inhomogeneous Euclidean group
and its central extension the Galilei group for the class of inertial
states in `nonrelativistic' quantum mechanics.\footnote{`Nonrelativistic'
means the\ \ approximation for velocities small relative to $c$.
The `nonrelativistic' has its own relativity group that is the Galilei
group in the inertial case. } 

In quantum mechanics, the obervables of position, time, energy and
momentum are the Hermitian representations of the algebra corresponding
to the unitary representation of the Weyl-Heisenberg group $\mathcal{H}(
n+1) $ \cite{weyl}\footnote{This is also often called the Heisenberg
group. It was Weyl who is generally credited with recognizing the
commutation relations was a Lie algebra and determining the group.
}\ \ It is the real matrix Lie group
\begin{equation}
\mathcal{H}( n+1) \simeq \mathcal{T}( n) \otimes _{s}\mathcal{T}(
n+1) .
\end{equation}

\noindent where $\mathcal{T}( n) \simeq (\mathbb{R}^{n},+)$. That
is, $\mathbb{R}^{n}$ considered to be a Lie group under addition.
The algebra has a basis $\{Z_{\alpha }\}=\{P_{i},Q_{i},E,T\}\in
a( \mathcal{H}( n+1) ) $, $i,j,..=1,..n$, and\ \ $\alpha ,\beta
,..=1,..2n+2$. The algebra has the familiar commutation relations.\ \ 
\begin{equation}
\left[ P_{i},Q_{j}\right] = \delta _{i,j}I,\ \ \ \ \ \ \left[ E,T\right]
=-I,
\end{equation}

The action of the group on the quantum states is given by the projective
representation $\pi $ that is the unitary representation $\varrho
$ of the central extension $\check{\mathcal{G}}$.\ \ \ The Weyl-Heisenberg\ \ is
a special case of the algebraic extension of the translation group
$\mathcal{T}( 2n+2) $.\ \ \ If $X_{\alpha }$ are the generators
of the abelian algebra of the translation group,\ \ then the central
extension is 
\begin{equation}
\left[ X_{\alpha },X_{\beta }\right] =M_{\alpha ,\beta },\ \ \left[
X_{\gamma },M_{\alpha ,\beta }\right] =0,\ \ M_{\alpha ,\beta }=-M_{\beta
,\alpha }\text{}%
\label{translation central extension candidates}
\end{equation}

The Weyl-Heisenberg group is the special case $M_{\alpha ,\beta
}=\zeta _{\alpha ,\beta }I$ where $\zeta _{\alpha ,\beta }$ is the
skew symmetric symplectic metric (0).\ \ Now as discussed in Appendix
A, the central extension of a group may be constrained if it is
the subgroup of a larger group.\ \ We know that position, time,
energy and momentum are correctly realized in the quantum mechanics
by the unitary representation of the Weyl-Heisenberg group and the
associated Hermitian representation of its algebra and so must be
constrained in this manner. Before turning to what this larger group
is, we briefly review this familiar Hermitian representation.\ \ 

The Mackey theorems for semidirect products (or the Stone von Neumann\footnote{The
Stone von Neumann theorem is specific to the Weyl-Heisenberg group
whereas the Mackey theorems apply to a general class of semidirect
product groups} theorem) may be used to compute the unitary irreducible
representations.\ \ The results are well known. The representations
are labelled by the eigenvalue of the single Casimir $I$,\ \ $\varrho
^{\prime }( I) |\psi \rangle =c|\psi \rangle $. The physical cases
is $c=\hbar $ for which the Hilbert space is $\text{\boldmath $\mathrm{H}$}\simeq
L^{2}( \mathbb{R}^{n+1},\mathbb{C}) $.\ \ \ If we choose a basis
$|q,t\rangle $ that\ \ diagonalizes ${\hat{Q}}_{i}$ and $\hat{T}$,
\begin{equation}
\begin{array}{cc}
 \left. \left. \left\langle  q,t\right. \right| {\hat{Q}}_{i}\left|
\psi \right. \right\rangle  =q^{i}\psi ( q,t) , & \left. \left.
\left\langle  q,t\right. \right| \hat{T}\left| \psi \right. \right\rangle
=-t \psi ( q,t) , \\
 \left. \left. \left\langle  q,t\right. \right|  {\hat{P}}_{i}\left|
\psi \right. \right\rangle  =i \hbar \frac{ \partial }{\partial
q^{i}}\psi ( q,t) , & \left. \left. \left\langle  q,t\right. \right|
\hat{E}\left| \psi \right. \right\rangle  =i \hbar \frac{ \partial
}{\partial  t}\psi ( q,t) , \\
 \left. \left. \left\langle  q,t\right. \right| \hat{I}\left| \psi
\right. \right\rangle  =\hbar  \psi ( q,t) . &  
\end{array}
\end{equation}

\noindent The representations satisfy the quantum commutation relations\footnote{The
$i$ is inserted in the exponential relating group and algebra\ \ so
that the algebra is represented by Hermitian rather than anti-Hermitian
operators and this results in its appearance in the commutation
relations.\ \ See [12].}\ \ \ . 
\begin{equation}
\left[ {\hat{P}}_{i},{\hat{Q}}_{j}\right] = i \hbar  \delta _{i,j},
\left[ \hat{E},\hat{T}\right] =-i \hbar ,%
\label{Heisenberg commutation relations representation}
\end{equation}

\noindent Of course, one could equally well choose the momentum
representation with basis $|p,t\rangle $ that diagonalizes ${\hat{P}}_{i}$
and $\hat{T}$ \cite{dirac}, or for that matter, a basis $|p,e\rangle
$ that diagonalizes ${\hat{P}}_{i},\hat{E}$\ \ or $|q,e\rangle $
that diagonalizes ${\hat{Q}}_{i},\hat{E}$.

We now return to the question of the larger group constraining the
algebraic central extension. A basic consistency condition exists
between the physical observables belonging to the algebra of the
Weyl-Heisenberg group and the relativity group that acts on them.\ \ We
consider the class of linear relativity groups that transform one
state into another so that in both states, position, time, energy
and momentum are represented by the Hermitian representation of
the Weyl-Heisenberg algebra. Each observers' specific measurements
of position, time, energy and momentum may differ, due to contractions
and dilations of the relativity transformation.\ \ However, in any
physical state that they are measured, they define a Weyl-Heisenberg
algebra.\ \ Then, from (0), for $g\in \check{\mathcal{G}}$, we have
\begin{equation}
\ \ \varrho ( g) {\hat{Z}}_{\alpha } \left| \psi \right\rangle 
=\varrho ( g) {\hat{Z}}_{\alpha } {\varrho ( g) }^{-1}\varrho (
g)  \left| \psi \right\rangle  ={{\hat{Z}}^{\prime }}_{\alpha }\left|
\psi ^{\prime }\right\rangle  ,%
\label{state transform}
\end{equation}

\noindent Therefore, as in (0-0), it follows that for a faithful
representation
\begin{equation}
{{\hat{Z}}^{\prime }}_{\alpha } =\varrho ^{\prime }( {Z^{\prime
}}_{\alpha }) = \varrho ( g) {\hat{Z}}_{\alpha }{\varrho ( g) }^{-1}=\varrho
^{\prime }( g Z g^{-1}) .%
\label{observable transform}
\end{equation}

\noindent and
\begin{equation}
{Z^{\prime }}_{\alpha } =g Z g^{-1},%
\label{generator transform}
\end{equation}

\noindent where we require that both ${Z^{\prime }}_{\alpha } ,Z_{\alpha
}\in \text{\boldmath $\mathrm{a}$}( \mathcal{H}( n+1) ) $.\ \ \ This
means that $g$ is an element of the automorphism group of the Heisenberg
algebra, ${\mathcal{A}ut}_{\mathcal{H}}$ and therefore the relativity
group $\mathcal{G}$ must be a subgroup, $\mathcal{G}\subset {\mathcal{A}ut}_{\mathcal{H}}$.\footnote{If
the observables are the Hermitian representation of the element
of the algebra of another group, then the same arguments hold and
the relativity group must be a subgroup of the automorphism group
of that algebra.} The automorphism group\ \ is given in \cite{folland}
and in Appendix A (0-0),
\begin{equation}
{\mathcal{A}ut}_{\mathcal{H}}\simeq \mathcal{A}\otimes _{s}\mathcal{Z}_{2}\otimes
_{s}\mathcal{H}\mathcal{S}p( 2n+2) ,%
\label{AutH}
\end{equation}

\noindent where
\begin{equation}
\mathcal{H}\mathcal{S}p( 2n+2) \simeq \mathcal{S}p( 2n+2) \otimes
_{s}\mathcal{H}( n+1) .%
\label{HSp original def}
\end{equation}

Therefore we have the result that, to be consistent with quantum
mechanics in which position, time, energy, and momentum are the
Hermitian representation of the Weyl-Heisenberg group, the relativity
group must be a subgroup of the automorphism group defined in (0-0).

This consistency may also be argued in the other direction. Suppose
the position, time, energy and momentum degrees of freedom are described
by the generators $\{Z_{\alpha }\}=\{P_{i},Q_{i},E,T\}$ that span
the algebra of $\mathcal{T}( 2n+2) $.\ \ This is the expected classical
description where the extended phase space is $\mathbb{R}^{2n+2}$.\ \ \ The
generators satisfy the commutation relations $[Z_{\alpha },Z_{\beta
}]=0$\ \ \ \ The quantum operators are the projective representations
of the generators, $\pi ^{\prime }( Z_{\alpha }) $, with $Z_{\alpha
}\in \text{\boldmath $\mathrm{a}$}( \mathcal{T}( 2n+2) ) $.\ \ The
projective representation is equivalent to the unitary representation
of the central extension of the group, $\varrho ^{\prime }( Z_{\alpha
}) $ with $X_{\alpha }\in \text{\boldmath $\mathrm{a}$}( \check{\mathcal{T}}(
2n+2) ) $.\ \ Considered as a group by itself, the algebraic extension
of the translation group is given in (0).

Suppose that a relativity group $\mathcal{G}$ relates the position,
time, energy and momentum Hermitian operators in different physical
states. The translation generators may be considered to be a subgroup
of the inhomogeneous group $\mathcal{I}\mathcal{G}=\mathcal{G}\otimes
_{s}\mathcal{T}( 2n+2) $.\ \ The group acting on the quantum states
is the Hermitian representation of the algebra of the central extension\ \ $\mathcal{I}\check{\mathcal{G}}$.\ \ If
we assume $\mathcal{G}\simeq \mathcal{S}p( 2n+2) $ then 
\begin{equation}
\mathcal{I}\mathcal{G}\simeq \mathcal{I}\mathcal{S}p( 2n+2) \simeq
\mathcal{S}p( 2n+2) \otimes _{s}\mathcal{T}( 2n+2) .%
\label{ISp original def}
\end{equation}

\noindent The central extension of the inhomogeneous symplectic
group is shown in Appendix B (0-0) to be $\mathcal{I}\check{\mathcal{S}}p(
2n+2)  =\mathcal{H}\mathcal{S}p( 2n+2) $ defined in (0).\ \ The
central elements for a central extension of the translation group
(0) have been constrained by its embedding in the inhomogeneous
symplectic group to be $M_{\alpha ,\beta }=\zeta _{\alpha ,\beta
}I$.

Therefore, in this case, the relativity group restricts the admissible
central extension of the translation group so that it is precisely
the Weyl-Heisenberg group. This turns out to be also true for certain
subgroups of the symplectic group $\mathcal{S}p( 2n+2) $ that we
shall turn to next.\ \ The system is quantized simply through the
projective representation that is required because physical states
are represented by rays in the Hilbert space. 

To summarize, consistency of the relativity group and quantum mechanics
requires that the relativity group is a subgroup of the group of
automorphisms of the\ \ Weyl-Heisenberg algebra of the basic physical
observables of position, time, energy and momentum.

On the other hand, we can start by assuming a relativity group that
is the symplectic group, or certain subgroups, and that physical
states are represented by rays in the Hilbert space with the corresponding
projective representations of the group. Then it directly follows
that the physical quantities of time, position, energy and momentum,
represented classically by translation generators, are the Hermitian
representations of the Weyl-Heisenberg algebra in the quantum formulation.
The {\itshape quantization} is directly a result of the states being
{\itshape rays} in a Hilbert space.
\section{Hamilton Relativity Group: Invariance of time}

Nonrelativistic quantum mechanics\ \ has the fundamental assumption
that time is an invariant for inertial and noninertial physical
states. All observers' clocks tick at the same rate.\ \ \ Time,
position, energy and momentum are represented by the generators
$Z_{\alpha }\in \text{\boldmath $\mathrm{a}$}( \mathcal{H}( n+1)
) $ where $\{Z_{\alpha }\}=\{P_{i},Q_{i},E,T\}$. The requirement
that a relativity group $\mathcal{G}$ leave $T$ invariant is, for
all $g\in \mathcal{G}$,
\begin{equation}
g T g^{-1}=T.
\end{equation}

\noindent This condition, together with the requirements that the
group is a subgroup of ${\mathcal{A}ut}_{\mathcal{H}}$,\ \ requires
that the $g$ be elements of the group 
\begin{equation}
\mathcal{H}\mathcal{S}p( n) \simeq \mathcal{Z}_{2}\otimes _{s}\mathcal{S}p(
2n) \otimes _{s}\mathcal{H}( n) .%
\label{HSP}
\end{equation}

\noindent This is established in Appendix A, (0,0) with the matrix
realization of $\mathcal{H}\mathcal{S}p( n) $ given in (0).

This identity is also established in \cite{Low7}\ \ and in that
reference it is shown that the diffeomorphisms of the extended phase
space\ \ $ \mathbb{P}$$ \simeq $$\mathbb{R}^{2n+2}$ into itself
whose Jacobians are elements of the group $\mathcal{H}\mathcal{S}p(
n) $ are Hamilton's equations.\ \ The Weyl-Heisenberg subgroup of
(0) is parameterized by the relative rate of change of\ \ position
$v^{i}$,\ \ momentum $f^{i}$ and energy $r$ with time. That is velocity,
force and power.\ \ Power is the central element.\ \ The generators
of velocity are $G_{i}$, force $F_{i}$ and power $R$.\footnote{Note
that the units of these are 1/velocity, 1/force and 1/power respectively}\ \ A
general element of this $\mathcal{H}( n) $\ \ algebra is\ \ 
\begin{equation}
v^{i}G_{i}+f^{i}F_{i}+r R.
\end{equation}

\noindent The full classical relativity group including the time,
position, energy and momentum generators is the group 
\begin{equation}
\mathcal{I}\mathcal{H}\mathcal{S}p( n) \simeq \mathcal{H}\mathcal{S}p(
n) \otimes _{s}\mathcal{T}( 2n+2) .%
\label{IHSP}
\end{equation}

For the action on the quantum physical states, we must determine
the unitary representation of the central extension.\ \ \,The most
general relativity group that has time as an invariant acting on
a quantum theory is therefore given by the unitary representations
of $\mathcal{I}\check{\mathcal{H}}\mathcal{S}p( n) $. The method
for computing central extensions is reviewed in Appendix A.\ \ It
can be shown that the algebraic central extension requires the addition
of the central element $I$ that turns the translation group $\mathcal{T}(
2n+2) $ in (0) into $\mathcal{H}( n+1) $.\ \ Another of the central
elements is mass and the third has the dimensions of reciprocal
tension as discussed in what follows. 

A further simplification is possible that makes the physical meaning
clearer by requiring invariance of the length $\delta ^{i,j}Q_{i}Q_{j}$
in the inertial rest frame. This eliminates substantial mathematical
complexity associated with non-orthonormal frames and enables the\ \ physical
meaning to be more transparent.\ \ The inertial rest frame action
of the group $\mathcal{H}\mathcal{S}p( n) $ in (0) has the parameters\ \ $v^{i}=f^{i}=r=0$
(see 0) and so we need only consider the symplectic subgroup.\ \ \ Thus
for\ \ $h\in \mathcal{S}p( 2n) $
\begin{equation}
\delta ^{i,j}Q_{i}Q_{j}= h \delta ^{i,j}Q_{i}Q_{j}h^{-1}.
\end{equation}

The required subgroup is $\mathcal{O}( n) \subset \mathcal{S}p(
2n) $\ \ (see \cite{Low7}\ \ and Appendix A) and it maps an orthonormal
basis $\{Q_{i}\}$ into an orthonormal basis $\{{\tilde{Q}}_{i}\}$\ \ This
defines the Hamilton group 
\begin{equation}
\mathcal{H}a( n) =\mathcal{Z}_{2}\otimes _{s}\mathcal{O}( n) \otimes
_{s}\mathcal{H}( n) \subset \mathcal{S}p( 2n) \otimes _{s}\mathcal{H}(
n) , 
\end{equation}

\noindent that is the relativity group for transformations between
frames in Hamilton's mechanics where the inertial rest position
frames are orthonormal. 

Again, the quantum theory requires us to consider\ \ the central
extension of 
\begin{equation}
\mathcal{I}\mathcal{H}a( n) \simeq \mathcal{H}a( n) \otimes _{s}\mathcal{T}(
2n+2) .
\end{equation}

\noindent As we will show in the next section, this central extension
is of the form\footnote{$\mathcal{Z}_{2}\otimes \mathcal{Z}_{2}$
is the 4 element parity, time reversal discrete group}
\begin{equation}
\begin{array}{rl}
 \mathcal{Q}\mathcal{H}a( n)  & =\mathcal{I}\check{\mathcal{H}}a(
n) \simeq \overline{\mathcal{H}a}( n) \otimes _{s}\left( \mathcal{T}(
2) \otimes \mathcal{H}( n+1) \right)  \\
  & \simeq \left( \mathcal{Z}_{2}\otimes \mathcal{Z}_{2}\right)
\otimes _{s}\overline{\mathcal{S}\mathcal{O}}( n) \otimes _{s}\mathcal{H}(
n) \otimes _{s}\left( \mathcal{T}( 2) \otimes \mathcal{H}( n+1)
\right) 
\end{array}%
\label{QHa}
\end{equation}

\noindent Again, the central extension of the translation group
in (0) is restricted by the inhomogeneous Hamilton relativity group
precisely so that it defines the Weyl-Heisenberg subgroup required
for the quantum realization of position, time, energy and momentum
as the Hermitian realization of Heisenberg generators. We will show
in the following section that it has 3 algebraic central generators.
Like the Galilei group, it has a nontrivial algebraic central extension
that is the mass generator $M$ and furthermore has a second central
element $A$ that has the physical dimensions that are the reciprocal
of tension (length/force) in addition to the central element $I$
that appears in the $\mathcal{H}( n+1) $ subgroup . $M$ and $A$
are generators of the algebra of the $\mathcal{T}( 2) $ translation
subgroup that appears in (0).
\section{Central Extension of the inhomogeneous Hamilton algebra}

The homogeneous and inhomogeneous Hamilton and Euclidean groups
as well as the Weyl-Heisenberg group are real matrix Lie groups.
The matrix realization of these groups is given in Appendix A.\ \ The
Lie algebras can therefore be directly computed from these matrix
realizations.\ \ The resulting nonzero commutation relations of
the algebra of $\mathcal{H}a( n) $ are
\begin{equation}
\begin{array}{c}
 \left[ J_{i,j},J_{k,l}\right] =J_{j,k} \delta _{i,l}+J_{i,l} \delta
_{j,k}-J_{i,k} \delta _{j,l}-J_{j,l} \delta _{i,k}, \\
 \left[ J_{i,j},G_{k}\right] =G_{j} \delta _{i,k}-G_{i}\delta _{j,k},
\\
 \left[ J_{i,j},F_{k}\right] =F_{i} \delta _{j,k}-F_{j}\delta _{i,k},
\\
 \left[ G_{i},F_{k}\right] =R \delta _{i,k}.
\end{array}%
\label{hamilton algebra}
\end{equation}

\noindent  The inhomogeneous Hamilton group $\mathcal{I}\mathcal{H}a(
n) $ requires the additional nonzero commutation relations: 
\begin{equation}
\begin{array}{cc}
 \left[ J_{i,j},Q_{k}\right] =-Q_{j} \delta _{i,k}+Q_{i} \delta
_{j,k}, & \left[ J_{i,j},P_{k}\right] =-P_{j} \delta _{i,k}+P_{i}
\delta _{j,k}, \\
 \left[ G_{i},Q_{k}\right] = \delta _{i,k}T, & \left[ F_{i},P_{k}\right]
= \delta _{i,k}T, \\
 \left[ E,G_{i}\right] =-P_{i}, & \left[ E,F_{i}\right] =Q_{i},
\\
 \left[ E,R\right] =2T. & \ \ 
\end{array}%
\label{inhomogeneous hamilton algebra}
\end{equation}

That $T$ is a central element as expected is clear from the structure
of the commutation relations.\ \ All physical states related by
group transformations generated by this algebra leave $T$ invariant.
All these states have the same definition of time.\ \ \ A general
element of the algebra is 
\begin{equation}
Z=\alpha ^{i,j}J_{i,j}+v^{i}G_{i}+f^{i}F_{i}+r R+ q^{i }P_{i}+t
E+p^{i}Q_{i}+e T.%
\label{general element of algebra}
\end{equation}

\noindent The $\alpha ^{i,j}$ are the $\frac{n( n-1) }{2}$ rotation
angles, $v^{i}$ velocity, $f^{i}$ force, $r$ power, $q^{i}$ position,
$t$ time, $p^{i}$ momentum and $e$ energy.\ \ \ Correspondingly
the generators have the dimensions such that $Z$ is dimensionless.
\subsection{Galilei group}

Before continuing with the Hamilton group, we briefly review the
inertial special case of the Hamilton group that is the familiar
Euclidean group of Galilean relativity \cite{Low7}.\ \ The Galilei
group is the central extension of the inhomogeneous Euclidean group
$\mathcal{I}\mathcal{E}( n) $ 
\begin{equation}
\mathcal{I}\mathcal{E}( n) \simeq \mathcal{E}( n) \otimes _{s}\mathcal{T}(
n+1) =\mathcal{S}\mathcal{O}( n) \otimes _{s}\mathcal{T}( n) \otimes
_{s}\mathcal{T}( n+1) .
\end{equation}

\noindent The algebra of\ \ $\mathcal{I}\mathcal{E}( n) \subset
\mathcal{I}\mathcal{H}a( n) $\ \ is spanned by the generators $\{J_{i,j},G_{i},P_{i},E\}$
that are a subset of the full set of generators in (0,0)
\begin{equation}
\begin{array}{c}
 \left[ J_{i,j},J_{k,l}\right] =J_{j,k} \delta _{i,l}+J_{i,l} \delta
_{j,k}-J_{i,k} \delta _{j,l}-J_{j,l} \delta _{i,k}, \\
 \left[ J_{i,j},G_{k}\right] =G_{j} \delta _{i,k}-G_{i}\delta _{j,k},
\\
 \left[ J_{i,j},P_{k}\right] =-P_{j} \delta _{i,k}+P_{i}\delta _{j,k},
\\
 \left[ E,G_{i}\right] = -P_{i}.
\end{array}%
\label{Galilei commutation relations}
\end{equation}

The central extension $\mathcal{G}a( n) \simeq \mathcal{I}\check{\mathcal{E}}(
n) $ may be directly computed using the method in Appendix A. It
is well known the algebraic central extension is the single generator
$M$ with the associated additional nonzero commutation relation\ \ 
\begin{equation}
\left[ G_{i},P_{k}\right] = \delta _{i,k}M.%
\label{Mass extension}
\end{equation}

\noindent Note that $M$ has the dimensions of mass and has a consistent
physical interpretation as mass. The Galilei group may be written
as
\begin{equation}
\mathcal{G}a( n) =\left( \mathcal{T}( 1) \otimes \overline{\mathcal{S}\mathcal{O}}(
n) \right) \otimes _{s}\mathcal{H}( n) .
\end{equation}

\noindent where $E$ is the generator of the algebra of the $\mathcal{T}(
1) $ group, $J_{i,j}$ are the generators of the algebra of $\overline{\mathcal{S}\mathcal{O}}(
n) $ and $\{G_{i},P{}_{i},M\}$ generate the Weyl-Heisenberg group
$\mathcal{H}( n) $ with $M$ the central element.\ \ Of course, in
the physical case $n=3$,\ \ $\overline{\mathcal{S}\mathcal{O}}(
3) =\mathcal{S}\mathcal{U}( 2) $.\ \ 
\subsection{Central extension of the inhomogeneous Hamilton group}

Returning to the central extension $\mathcal{I}\check{\mathcal{H}}a(
n) $ of the Hamilton group, direct symbolic Lie algebra computation
using {\itshape Mathematica} with the method described in Appendix
A results in the addition of the central elements\ \ 
\begin{equation}
\left[ P_{i},Q_{k}\right] = \delta _{i,k}I,\ \ \left[ E,T\right]
=-I,\ \ \ \left[ G_{i},P_{k}\right] = \delta _{i,k}M,\ \ \left[
F_{i},Q_{k}\right] = \delta _{i,k}A,%
\label{AIM central extension of IHa}
\end{equation}

\noindent to the Hamilton algebra defined in (0,0). The three new
central elements $\{M,A,I\}$, ($N_{e}=3$), result in the following
terms being added to a general element of the algebra given in (0)
\begin{equation}
Z=\alpha ^{i,j}J_{i,j}+v^{i}G_{i}+f^{i}F_{i}+r R+ q^{i }P_{i}+t
E+p^{i}Q_{i}+e T +a A+m M+\iota  I.%
\label{IHa central extension general algebra element}
\end{equation}

The central extension condition for $M$ in (0) is identical to the
mass central extension in the Galilei group (0) and is precisely
the condition for $\mathcal{G}a( n) $ to be the inertial subgroup
of\ \ $\mathcal{Q}\mathcal{H}a( n) $.\ \ The central extension $I$,
with dimensions of action, is precisely the condition for the unitary
representation to yield the usual Heisenberg commutation relations.\ \ \ \ The
final extension $A$ is new and has the dimensions that is the reciprocal
of tension,\ \ length/force.\ \ Therefore the parameter $a$ has
dimensions of tension, $m$ of reciprocal mass and $\mathrm{\iota
}$ of reciprocal action.\ \ 
\section{Casimir invariants }

Casimir invariant operators $C_{\alpha }$, $\alpha =1,..N_{c}$\ \ are
polynomials in the enveloping algebra that commute with all the
generators of the algebra of the group in question $[C_{\alpha },Z_{A}]=0$.\ \ \ $Z{}_{A}\in
\text{\boldmath $\mathrm{a}$}( \mathcal{G}) $, $A=1,...N_{g}$ and
$N_{g}$ is the dimension of $\mathcal{G}$. $\alpha =1,..N_{c}$ where
$N_{c}$ is the number of Casimir invariants.\ \ The eigenvalues
$\nu _{\alpha }$ of the Hermitian representations of the Casimir
invariants ${\hat{C}}_{\alpha }|\psi \rangle =\nu _{\alpha }|\psi
\rangle $, $\nu _{\alpha }\in \mathbb{R}$ are invariants for all
physical states related by the relativity group $\mathcal{G}$.\ \ These
invariants typically label irreducible unitary representations (but
not always completely) and represent fundamental physical quantities.
For example, in both the Galilei\ \ and the Poincar\'e relativity
group, mass and spin are the eigenvalues of the representations
of the corresponding Casimir invariants.\ \ 

The number $N_{c}$ of Casimir invariants may be computed directly
from a theorem that states that it is $N_{c}=N_{g}-N_{r}$. $N_{r}$
is the rank of the $N_{g}\times N_{g}$ matrix $z^{A}c_{A,B}^{C}$
that is the adjoint representation of the algebra\ \ $[Z_{A},Z_{B}]=c_{A,B}^{C}Z_{C}$
\cite{Gilmore}. A general element of the algebra is $Z=z^{A}Z_{A}$.\ \ The
Casimirs can be found by constructing a general element $p^{l}(
Z_{A}) $ of the enveloping algebra up to a give order of polynomial
in the generators with general coefficients.\ \ Setting $[p^{l}(
Z_{A}) ,Z_{A}]=0$ creates a linear set of equations in the coefficients
that can be solved to determine the Casimir invariant operators.
A central element is a polynomial or order 1 and so the first $N_{e}$
Casimir invariants are the central elements. This is a conceptually
a simple calculation\ \ but is best carried out using a symbolic
computation package written in {\itshape Mathematica }\cite{wolfram}
as the number of bracket computations and the linear equations is
large for the algebras in question.\ \ \ \ 

The results for the dimensions are given in the following table.
Of course, while $N_{g}$ may be determined for general $n$ and $N_{e}$
is independent of $n$, the $N_{c}$ must be computed from the rank
of the symbolic matrix on a case by case basis.\ \ For the remainder
of this section we will restrict our attention to $n\leq 3$.
\begin{equation}
\begin{array}{cccccc}
   & N_{g} & N_{e} & N_{c}( n=1)  & N_{c}( n=2)  & N_{c}( n=3) 
\\
 \mathcal{G}a( n) : & \frac{1}{2} \left( n^{2}+3 n+4\right)  & 1
& 2 & 3 & 3 \\
 \mathcal{I}\check{\mathcal{H}}a( n) : & \frac{1}{2} \left( n^{2}+7
n+12\right)  & 3 & 4 & 5 & 5
\end{array}%
\label{Ng Ne Nc}
\end{equation}

Before considering the group $\mathcal{Q}\mathcal{H}a( n) =\mathcal{I}\check{\mathcal{H}}a(
n) $ we briefly review the well known results from the Galilei group,
$\mathcal{G}a( n) \simeq \mathcal{I}\check{\mathcal{E}}( n) $.\ \ For
$n\leq 3$, there are 3 Casimir invariants, one of which is the central
element $M$,
\begin{equation}
\begin{array}{cc}
 C_{1}=M, & C_{2}=2M E-P_{i} P_{i}, \\
 C_{3}=M^{2 }S_{i,j} S_{i,j}, & S_{i,j}=J_{i,j}-\frac{1}{M}\left(
G_{j}P_{i}-G_{i}P_{j}\right) .
\end{array}%
\label{Galilei group Casimirs}
\end{equation}

\noindent Note that $C_{3}$ is identically zero for $n=1$.\ \ From
(0),\ \ it\ \ follows that these are the complete set of Casimir
invariants for\ \ $\mathcal{G}a( n) $ for $n\leq 3$.\ \ For clearer
physical insight, note that the Casimir invariants $C_{2}$ and $C_{3}$
can be written as the energy and spin of an inertial (free) particle
\begin{equation}
 E-E \mbox{}^{\circ}=\frac{1}{2M}P_{i}P_{i}\ \ \ , S^{2}=S_{i,j}
S_{i,j}\ \ \mathrm{where}\ \ \ E \mbox{}^{\circ}=\frac{1}{2M}C_{2}.%
\label{spin energy Casimir}
\end{equation}

The Casimir invariants of\ \ $\mathcal{Q}\mathcal{H}a( n) $ may
be directly computed using the same method. For $n\leq 3$, there
are 5 Casimir invariants, 3 of which are the central elements $I,M,A$,
\begin{equation}
\begin{array}{c}
 C_{1}=I, C_{2}=M, C_{3}=A, \\
 C_{4}=T T-I R, \\
 C_{5}=C^{2}B_{i,j}B_{i,j},
\end{array}%
\label{Hamilton casimirs}
\end{equation}

\noindent where
\begin{equation}
\begin{array}{c}
 C=C_{2} C_{3}+C_{4}=-A M+T^{2} -I R , \\
 B_{i,j}=J_{i,j}+\frac{1}{C}D_{i,j}.
\end{array}%
\label{noninertial spin}
\end{equation}

\noindent  $C$ is a Casimir invariant as any polynomial combinations
of a Casimir is a Casimir and the $D_{i,j}$ are given by
\begin{equation}
D_{i,j}=A D_{i,j}^{1}+M D_{i,j}^{2}+R D_{i,j}^{3}+ I D_{i,j}^{4}+T
\left( D_{i,j}^{5}+D_{i,j}^{6}\right) ,%
\label{noninertial spin components}
\end{equation}

\noindent where
\begin{equation}
\begin{array}{cc}
 D_{i,j}^{1}=G_{j} P_{i}-G_{i} P_{j}, & D_{i,j}^{2}=F_{j} Q_{i}-F_{i}
Q_{j}, \\
 D_{i,j}^{3}=P_{i} Q_{j}-P_{j} Q_{i}, & D_{i,j}^{4}=F_{i} G_{j}-F_{j}
G_{i}, \\
 D_{i,j}^{5}=F_{i} P_{j}-F_{j} P_{i}, & D_{i,j}^{6}=G_{i} Q_{j}-G_{j}
Q_{i}.
\end{array}
\end{equation}

\noindent $B_{i,j}$ vanishes for $n=1$. It is a straightforward
computation using the Lie algebra relations (0,0,0) to verify that
these are invariant.\ \ From (0),\ \ it\ \ follows that these are
the complete set of Casimir invariants for\ \ $\mathcal{Q}\mathcal{H}a(
n) $ for $n\leq 3$.\ \ Note that $B_{i,j}$ may also be written as
$B_{i,j}=S_{i,j}+\frac{1}{C} {\tilde{D}}_{i,j}$ where $S_{i,j}$
is the Galilean spin defined in (0) and\ \ as $\frac{1}{M}+\frac{A}{C}=\frac{C_{4}}{M
C}$
\begin{equation}
{\tilde{D}}_{i,j}=\frac{C_{4}}{M} D_{i,j}^{1}+M D_{i,j}^{2}+R D_{i,j}^{3}+
I D_{i,j}^{4}+T \left( D_{i,j}^{5}+D_{i,j}^{6}\right) 
\end{equation}

The eigenvalues of the Casimirs in the unitary representation usually
label irreducible representations. In a representation where the
eigenvalue of $C$ goes to zero, the $D_{i,j}$ term will be negligible
and the spin reduces to the usual Galilean spin $\operatorname*{\lim
}\limits_{C\rightarrow \:0}B_{i,j} =S_{i,j}.$ A sufficient condition
for this is $C_{3}=A\rightarrow 0$ and $C_{4}\rightarrow 0$.\ \ As
$A=0$ means that the tension $1/A$ is infinite. 

\noindent 
\section{Discussion}

Physical states in quantum mechanics are represented by rays in
a Hilbert space. Quantum mechanics realizes position, time, momentum
and energy as the Hermitian representation of the generators ${\hat{Z}}_{\alpha
}$ of the algebra of the Weyl-Heisenberg group. These generators
acting on a state, ${\hat{Z}}_{\alpha }|\psi \rangle $, are transformed
by the unitary representations of a relativity group to define generators
acting on the transformed state, ${\hat{\tilde{Z}}}_{\alpha }\tilde{|\psi
\rangle }$.\ \ In order for the transformed generators to also be
generators of the Weyl-Heisenberg group, the relativity group must
be a subgroup of the group of automorphisms of the Weyl-Heisenberg
algebra. If the relativity group does not have this property, then
the position, time, momentum and energy degrees of freedom would
not satisfy the Heisenberg commutation relations in the transformed
state, as given in (0).\ \ This provides a basic consistency condition
between the relativity group and the Weyl-Heisenberg group of quantum
mechanics.

Folland has proven that the automorphism group of the Weyl-Heisenberg
algebra (and group) is the group $\mathcal{H}\mathcal{S}p( 2n+2)
$ together with a conformal scaling group $\mathcal{A}$ and a two
element discrete group $\mathcal{Z}_{2}$ that reverses the sign
of time and energy.\ \ \ The Weyl-Heisenberg subgroup of $\mathcal{H}\mathcal{S}p(
2n+2) $ are the inner automorphisms. The\ \ constraint on the continuous
homogeneous relativity group is that it is a subgroup of the symplectic
subgroup and conformal scaling group.\ \ This shows the very deep
connection between the Weyl-Heisenberg group and the symplectic
structure.\ \ The symplectic structure does not need to be postulated
{\itshape a priori} but is simply required by this consistency with
quantum mechanics. 

Invariance of classical time results in the $\mathcal{H}\mathcal{S}p(
2n) $ homogeneous relativity group that is a subgroup of $\mathcal{S}p(
2n+2) $.\ \ We have previously shown that the requirement that Jacobians
of diffeomorphisms of classical extended phase space into itself
be elements of this group are Hamilton's equations.\ \ While this
is the most general group, a considerable amount of the mathematical
generality of this group is to accommodate frames that are not orthonormal
in the inertial rest frame. Requiring invariance of length in the
inertial rest frame reduces this group to the physical Hamilton
group $\mathcal{H}a( n) $ \cite{Low7}.\ \ 

The central generators of mass, fundamental to classical mechanics,
and the $I$ required to give the Heisenberg commutation relations
in the representation appear automatically as a result of the projective
representations that are required because the physical quantum states
are rays in a Hilbert space.

Projective representations of this group act on quantum states that
are generally noninertial. Here we get our first surprise. In the
quantum realization,\ \ time is not an invariant. Rather, taking
the representation of\ \ (0) we obtain an expression of the form
acting on a quantum state
\begin{equation}
T^{2} \left| \psi \right\rangle  =\left( \tau ^{2}+ \hbar  R\right)
\left| \psi \right\rangle  ,
\end{equation}

\noindent where $\tau ^{2}$ is the eigenvalue of\ \ the Casimir
$C_{4}$.\ \ 

The next interesting fact is that there is no Casimir operator (such
as $C_{2}$ for the Galilei group) that involves the energy generator.\ \ \ However,
when one recalls from basic classical mechanics that noninertial
frames do not conserve energy, this is not so surprising and to
be expected.\ \ What is surprising is that there is a natural generalization
of spin that is invariant in all noninertial physical states.\ \ All
observers, inertial and noninertial,\ \ calculate the same values
for the eigenvalues of\ \ the Casimir $C_{5}$ that may be expressed
as Galilei spin with additional terms (0,0).\ \ This provides a
possibility of testing this theory by studying spin in noninertial
states in `nonrelativistic' quantum mechanics.\ \ 

Finally, an even more surprising result is the appearance of a new
central generator $A$. The parameter $a$ for the term $a A$ in the
general term for the algebra (0)\ \ has the dimensions of tension.\ \ This
generator shows up in the algebra in as fundamental a manner as
the mass and $I$ generator. Both of those are rather basic to physics
and so a critical test of these ideas is a further physical understanding
of $A$.\ \ Note $A$ is reciprocally dual to $M$ in the sense of
Born \cite{born2}. This reciprocal symmetry underlies the material
presented. 

As noted, the eigenvalues of the Hermitian representations of the
Casimir invariants for the unitary representations of the group
typically label the irreducible representations.\ \ There will probably
be a class of irreducible representations with states where $\hat{A}|\psi
\rangle =0$.\ \ In these states {\itshape tension} is infinite.
Tension seems to play a basic role in string theory and so it is
intriguing that a tension term shows up here. 

The unitary representations of the group $\mathcal{Q}\mathcal{H}a(
n) $, and many of the other groups in this paper, have a rich semidirect
product structure. The methods of Mackey for determining the unitary
representations of semidirect product groups can be used to determine
the representations \cite{Low}. This will be undertaken in a subsequent
paper to complete the quantum description of this relativity group.

The same methods may be generalized to relativity groups with other
invariants. For example, instead of the invariant $T$, we could
have the Minkowski invariant $T^{2}-\frac{1}{c^{2}}Q^{2}$ appropriate
for special relativity or the Born line element $T^{2}-\frac{1}{c^{2}}Q^{2}-\frac{1}{b^{2}}P^{2}+\frac{1}{c^{2}b^{2}}E^{2}$
appropriate for reciprocal relativity \cite{Low6}.\footnote{We note
that $\mathcal{Q}\mathcal{H}a( n) $ is the $b,c\rightarrow \infty
$ limit of the quaplectic group that appears in reciprocal relativity.}\ \ \ In
particular, one would conjecture a generalization of the standard
spin in relativistic quantum mechanics that involves additional
terms similar to what has been shown here in the classical context
required for noninertial frames to provide a critical test. 

I would like to thank Peter Jarvis for many discussions that have
clarified the ideas presented here. 

\appendix

\section{Central Extensions}\label{Appendix central extension}

The central extension $\check{\mathcal{G}}$ of a group $\mathcal{G}$
is defined by the short exact sequence
\[
1 \rightarrow \mathcal{A}\rightarrow \check{\mathcal{G}}\rightarrow
\mathcal{G}\rightarrow 1
\]

\noindent where $ \mathcal{A}$ is an abelian group in the center
of $\check{\mathcal{G}}$ that is the extension.\ \ This induces
a central extension for the algebra of $\check{\mathcal{G}}$ corresponding
to the algebra of $\mathcal{A}$.\ \ The set of isomorphism classes
of the central extension of $\mathcal{G}$ by $ \mathcal{A}$ is in
one to one correspondence with the second cohomology group $H^{2}(
\mathcal{G},\mathcal{A}) $. The methods of algebraic topology that
may be used to determine this cohomology group are described in
\cite{Azcarraga}.\ \ \ 

Alternatively, a central extension $\check{\mathcal{G}}$ is the
universal cover of the group whose algebra is the central extension
of the algebra of\ \ $\mathcal{G}$.\ \ The central extension of
the algebra is explicitly constructed as the most general central
extension satisfying the Jacobi identities for the algebra as described
in \cite{Weinberg1}.\ \ A nontrivial first homotopy group for the
group $ \mathcal{G}$ results in the topological extension that is
the first homotopy group in the construction of the universal cover.
We refer to the extension of the algebra as the algebraic extension
and the cover as the topological extension.\ \ This indirectly results
in the cohomology group $H^{2}( \mathcal{G},\mathcal{A}) $ due to
the equivalence noted.\ \ This method of determining the algebraic
extension is tractable by creating a set of general Lie algebra
evaluation rules in {\itshape Mathematica }\cite{wolfram}

Consider a general Lie algebra with basis $Z_{\alpha }\in \text{\boldmath
$\mathrm{a}$}( \mathcal{G}) $, $\alpha ,\beta ,..=1,..N_{g}$, with
$N_{g}$ the dimension of the group and algebra,\ \ satisfying commutation
relations 
\begin{equation}
\left[ Z_{\alpha },Z_{\beta }\right] =c_{\alpha ,\beta }^{\gamma
}Z_{\gamma }.%
\label{General Lie Algebra}
\end{equation}

\noindent A general element of the algebra is $Z=z^{\alpha }Z_{\alpha
}$\ \ with $z^{\alpha }\in \mathbb{R}$.\ \ The central extension
of the algebra\ \ is defined by the maximal addition of central
elements $M_{\alpha ,\beta }$ to the algebra\ \ that are consistent
with the Jacobi identities.\ \ First, construct the maximal set
of candidate extensions as
\begin{equation}
\left[ Z_{\alpha },Z_{\beta }\right] =c_{\alpha ,\beta }^{\gamma
}Z_{\gamma }+M_{\alpha ,\beta },\ \ \left[ Z_{\gamma },M_{\alpha
,\beta }\right] =0.%
\label{central extension candidate}
\end{equation}

The problem of determining the central extension is to find the
most general set of $M_{\alpha ,\beta }$ for which the Jacobi identities
for the set of generators are satisfied
\begin{equation}
\left. \left[ Z_{\alpha },\left[ Z_{\beta },Z_{\gamma }\right] \right]
+\left[ Z_{\beta },\left[ Z_{\gamma }\right] ,Z_{\alpha }\right]
\right] =\left[ Z_{\gamma },\left[ Z_{\alpha },Z_{\beta }\right]
\right] =0,%
\label{Jacobi Identities}
\end{equation}

\noindent Clearly the combination $M_{\alpha ,\beta }=c_{\alpha
,\beta }^{\gamma }M_{\gamma }$ is always a solution constituting
to translating each generator by a corresponding central element,\ \ ${\check{Z}}_{\alpha
}=Z_{\alpha }+M_{\alpha }$. These are discarded as trivial. The
remaining solutions define the central extension.

Consider a Lie group $\mathcal{G}$ with a candidate central extension
of its\ \ algebra given by (0). The number of Jacobi identities
can be dramatically reduced by examining the properties of the central
extensions for subsets of the generators $\{Z_{a}\}$ that define
subalgebras. A necessary condition that the full algebra admit a
central extension is that the subalgebra also admit the extension.\ \ If
the subalgebra does not have an extension, we can immediately set
the corresponding set of candidate central elements to zero $\{M_{a,b}=0\}$.\ \ \ The
embedding in the larger group in these cases blocks the extension
from being an extension of the full group.\ \ \ 

It is well known that the following groups do not have an algebraic
central extension: $\mathcal{O}( n) $, $\mathcal{O}( 1,n) $, $\mathcal{S}p(
2n) $, $\mathcal{E}( n) $, $\mathcal{E}( 1,n) $. (See for example
\cite{Weinberg1} for a detailed analysis of $\mathcal{E}( 1,n) $.)\ \ $\mathcal{E}(
n) =\mathcal{S}\mathcal{O}( n) \otimes _{s}\mathcal{T}( n) $ provides
an immediate example where the subgroup $\mathcal{T}( n) $ has a
central extension that is blocked by the fact that it is a subgroup
of the semidirect product with $\mathcal{S}\mathcal{O}( n) $ as
the homogeneous group. 

The algebraic central extension of the inhomogeneous symplectic
group may be calculated in the same manner.\ \ Suppose $\{W_{a,b},
Y_{a}\}$ are the generators of the algebra of the inhomogeneous
symplectic group,\ \ $\mathcal{I}\mathcal{S}p( 2n+2) \simeq \mathcal{S}p(
2n+2) \otimes _{s}\mathcal{T}( 2n) $. The nonzero commutation relations
are
\begin{gather}
\left[ W_{\alpha ,\beta },W_{\kappa ,\delta }\right] =\zeta _{\beta
,\kappa }W_{\alpha ,\delta }+\zeta _{\alpha ,\kappa }W_{\beta ,\delta
}+\zeta _{\beta ,\delta }W_{\alpha ,\kappa }+\zeta _{\alpha ,\delta
}W_{\beta ,\kappa }%
\label{ISp Lie algebra}
\\\left[ W_{\alpha ,\beta },Y_{\kappa }\right] =\zeta _{\beta ,\kappa
}Y_{\kappa }+\zeta _{\alpha ,\kappa }Y_{\kappa }
\end{gather}

\noindent where $\alpha ,\beta ,..=1,...2n+2$.\ \ \ Immediately
$M_{\alpha ,\beta ,\kappa ,\delta }^{W,W}=0$ as $\mathcal{S}p( 2n+2)
$ does not admit a central extension. It is then simply a matter
of introducing the candidate extensions $\{M_{\alpha ,\beta }^{Y,Y},M_{\alpha
,\beta ,\kappa }^{W,Y}\}=\{M_{\alpha ,\beta },M_{\alpha ,\beta ,\kappa
}\}$ and checking the Jacobi relations when they are added to the
above generators and the implicit relation 
\begin{equation}
\left[ Y_{\alpha },Y_{\beta }\right] =M_{\alpha ,\beta }
\end{equation}

\noindent The essential Jacobi relations from (0) are 
\begin{equation}
\begin{array}{c}
 \left\{ Y_{\alpha },W_{\kappa ,\delta },W_{\gamma ,\epsilon }\right\}
=\zeta _{\epsilon ,\kappa } M_{\gamma ,\delta ,\alpha }+\zeta _{\alpha
,\kappa } M_{\gamma ,\epsilon ,\delta }+\zeta _{\alpha ,\delta }
M_{\gamma ,\epsilon ,\kappa } \\
 -\zeta _{\delta ,\epsilon } M_{\gamma ,\kappa ,\alpha }+\zeta _{\gamma
,\kappa } M_{\delta ,\epsilon ,\alpha }-\zeta _{\alpha ,\epsilon
} M_{\delta ,\kappa ,\gamma }-\zeta _{\alpha ,\gamma } M_{\delta
,\kappa ,\epsilon }+\zeta _{\gamma ,\delta } M_{\epsilon ,\kappa
,\alpha } \\
 \left\{ Y_{\alpha },Y_{\kappa },W_{\gamma ,\epsilon }\right\} =-M_{\kappa
,\epsilon } \zeta _{\alpha ,\gamma }-M_{\kappa ,\gamma } \zeta _{\alpha
,\epsilon }-M_{\alpha ,\epsilon } \zeta _{\gamma ,\kappa }-M_{\alpha
,\gamma } \zeta _{\epsilon ,\kappa }
\end{array}%
\label{ISp Jacobi Identities}
\end{equation}

It can be verified that $M_{\gamma ,\delta ,\alpha }$ only has trivial
solutions where-as $M_{\alpha ,\beta }=\zeta _{\alpha ,\beta }I$
where $I$ is a central element is a nontrivial solution.\ \ Thus,
the algebraic central extension is the group $\mathcal{H}\mathcal{S}p(
n) \simeq \mathcal{S}p( 2n) \otimes _{s}\mathcal{H}( n) $ as claimed.

As another example, this method can be immediately used to reduce
the number of central generators that need to be checked with the
Jacobi conditions for the Galilei group.\ \ Here the candidate central
elements (with superscripts labeling the commutators to which they
apply)\ \ are \{$M_{i,j,k,l}^{J,J}$, $M_{i,j,k}^{J,G}$, $M_{i,j,k}^{J,P}$,\ \ $M_{i,j}^{J,E}\text{}$,\ \ $M_{i}^{G,E}$,
$M_{i}^{P,E}$, $M_{i,j}^{G,P}$\}.\ \ \ As $\{J_{i,j},G_{i}\}$ and
$\{J_{i,j},P_{i}\}$ are the generators of Euclidean subalgebras,
we can immediately set $M_{i,j,k,l}^{J,J}=0$, $M_{i,j,k}^{J,G}=0$,
$M_{i,j,k}^{J,P}=0$ significantly reducing the number of Jacobi
identities that need to be calculated. 

The same is true of the inhomogeneous\ \ Hamilton group where the
relations\ \ \{$M_{i,j,k}^{J,F}$,\ \ $M_{i,j,k}^{J,Q}$\} may also
be set to zero as they are the generators of\ \ Euclidean subalgebras.\ \ The
remainder need to be checked directly through the Jacobi identities
that is best undertaken using the symbolic computation capabilities
of {\itshape Mathematica.} 

\section{Matrix realizations of the groups}\label{Appendix Matrix
group}

The homogeneous and inhomogeneous Hamilton and Euclidean groups
and the Weyl-Heisenberg groups may be realized as matrix groups
that are subgroups of $\mathcal{G}\mathcal{L}( 2n+2,\mathbb{R})
$.\ \ Elements of this group are nonsingular $(2n+2)\times (2n+2)$
real matrices.\ \ 

The group element\ \ $\Gamma \mbox{}^{\circ}( \epsilon ,A,w,\iota
) \in \mathcal{H}\mathcal{S}p( 2n) $\ \ and Hamilton group $\Phi
\mbox{}^{\circ}( \epsilon ,R,v,f,\iota ) \in \mathcal{H}a( n) $
are matrix subgroups of $\mathcal{G}\mathcal{L}( 2n+2,\mathbb{R})
$ \cite{Major,Low7} with the form
\begin{equation}
\Gamma \mbox{}^{\circ}=\left( \begin{array}{ccc}
 A & 0 & w \\
 {}-^{t}w \zeta \mbox{}^{\circ} A & \epsilon  & r \\
 0 & 0 & \epsilon 
\end{array}\right)  ,\ \ \Phi \mbox{}^{\circ}=\left( \begin{array}{cccc}
 R & 0 & 0 & f \\
 0 & R & 0 & v \\
 {}^{t}v R & -{}^{t}f R & \epsilon  & r \\
 0 & 0 & 0 & \epsilon 
\end{array}\right) \text{},%
\label{matrix reps of HSp and Ha}
\end{equation}

\noindent where $r\in \mathbb{R}$, $\epsilon =\pm 1$\ \ \ and\ \ $w\in
\mathbb{R}^{2n}$, , $\epsilon =\pm 1$, $A\in \mathcal{S}p( 2n) $
realized by $2n \times 2n$ matrices in $\Gamma \mbox{}^{\circ}$\ \ and
$f,v\in \mathbb{R}^{n}$,\ \ $R\in \mathcal{O}( n) $ realized by
$n \times n$ matrices in $\Phi \mbox{}^{\circ}$. The symplectic
metric is
\begin{equation}
\zeta \mbox{}^{\circ}=\left( \begin{array}{cc}
 0 & I_{n} \\
 -I_{n} & 0
\end{array}\right)  ,%
\label{symplectic matrix}
\end{equation}

\noindent $I_{n}\ \ $$\mathrm{is} \mathrm{the} n \times n \mathrm{unit}
\mathrm{matrix}$. The subgroup chain 
\begin{equation}
\mathcal{H}( n) \subset \mathcal{H}a( n) \subset \mathcal{H}\mathcal{S}p(
2n) \subset \mathcal{G}\mathcal{L}( 2n+2,\mathbb{R}) ,
\end{equation}

\noindent leads to the identifications of $w=(f,v)$ and also note
that $\mathcal{O}( n) \subset \mathcal{S}p( 2n) $.\ \ 

Elements of the Weyl-Heisenberg group are given by either $\Upsilon
( w,\iota ) =\Gamma \mbox{}^{\circ}( 1,I_{2n},w,\iota ) $ or $\Upsilon
( v,f,\iota ) =\Phi \mbox{}^{\circ}( 1,I_{n},v,f,\iota ) $.\ \ The
group multiplication, inverses and automorphisms may be computed
simply through matrix multiplication and inverse. The basis of the
Lie algebra is give by differentiating the matrices by the parameters
and evaluating at the identity. The Lie algebra structure relations
are computed directly by matrix multiplication to establish the
abstract relations in (0). 

Likewise, for the inhomogeneous group, we have the inclusion chain.
\begin{equation}
\mathcal{I}\mathcal{E}( n) \subset \mathcal{I}\mathcal{H}a( n) \subset
\mathcal{I}\mathcal{H}\mathcal{S}p( 2n+2) \subset \mathcal{I}\mathcal{G}\mathcal{L}(
2n+2,\mathbb{R}) \subset \mathcal{G}\mathcal{L}( 2n+3,\mathbb{R})
.
\end{equation}

\noindent The corresponding matrix representations of the inhomogeneous
groups $\Gamma ( \epsilon ,A,w,r,z,e,\iota ) \in \mathcal{I}\mathcal{H}\mathcal{S}p(
2n+2) $\ \ and\ \ $\Phi ( \epsilon ,R,v,f,r,q,p,e,t) $ $\in \mathcal{I}\mathcal{H}a(
n) $ are 
\begin{equation}
 \Gamma =\left( \begin{array}{cccc}
 A & 0 & w & z \\
 -^{t}w \zeta \mbox{}^{\circ} A & \epsilon  & r & e \\
 0 & 0 & \epsilon  & t \\
 0 & 0 & 0 & 1
\end{array}\right)  ,\ \ \ \Phi =\left( \begin{array}{ccccc}
 R & 0 & 0 & f & p \\
 0 & R & 0 & v & q \\
 {}^{t}v R & -{}^{t}f R & \epsilon  & r & e \\
 0 & 0 & 0 & \epsilon  & t \\
 0 & 0 & 0 & 0 & 1
\end{array}\right) \text{}.%
\label{matrix realization of IHSp}
\end{equation}

In these expressions $z\in \mathbb{R}^{2n}$, $p,q\in \mathbb{R}^{n}$
and there is the identification $z=(p,q)$.\ \ Elements of the form
$\Phi ( \epsilon ,R,v,0,0,p,0,e,0) $ define an inhomogeneous Euclidean
subgroup $\mathcal{I}\mathcal{E}( n) $, the central extension of
which is the Galilei group.\ \ 

Finally $\mathcal{H}\mathcal{S}p( n) \simeq \mathcal{S}p( 2n+2)
\otimes _{s}\mathcal{H}( n) $ is a matrix subgroup of $\mathcal{G}\mathcal{L}(
2n+4,\mathbb{R}) $.\ \ 
\begin{equation}
\Phi ( \epsilon ,A,w,\iota ,z,e,t) =\left( \begin{array}{ccccc}
 A & 0 & w & 0 & z \\
 -^{t}w \zeta \mbox{}^{\circ} A & \epsilon  & \iota  & 0 & e \\
 0 & 0 & \epsilon  & 0 & t \\
 -^{t}z \zeta \mbox{}^{\circ} A & -t & e & 1 & \iota  \\
 0 & 0 & 0 & 0 & 1
\end{array}\right)  .\ \ 
\end{equation}

\noindent The central extension of a matrix group is not necessarily
a matrix group \cite{Hall}.\ \ It is not been established whether
the extensions $\check{\mathcal{H}}\mathcal{S}p( n) $ and $\mathcal{Q}\mathcal{H}a(
n) $ are matrix groups.
\subsection{Automorphisms of the Weyl-Heisenberg Group}\label{Appendix
Matrix group}

${\mathcal{A}ut}_{\mathcal{H}}$ that is given by (0) is proven by
Folland. (See page 20 of \cite{folland} We provide here the matrix
representation and group composition law and explicitly compute
the automorphisms.\ \ Using the definition $\Upsilon ( w,\iota )
=\Gamma ( I_{2n},w,\iota ) $ and (0) the Weyl-Heisenberg group $\mathcal{H}(
n) $ composition law is the expected
\begin{equation}
\Upsilon ( w^{\prime },\iota ^{\prime }) \cdot \Upsilon ( w,\iota
) =\Upsilon ( w+w^{\prime },\iota +\iota ^{\prime }+w^{\prime }\cdot
\zeta  \mbox{}^{\circ}\cdot w) ,\ \ {\Upsilon ( w,\iota ) }^{-1}=\Upsilon
( -w,-\iota ) .
\end{equation}

Elements of the linear automorphism group ${\mathcal{A}ut}_{\mathcal{H}}$
may be represented by $(2n+2)\times (2n+2)$ matrices $\Omega $ that
satisfy $\Omega  \Upsilon ( w^{\prime },\iota ^{\prime })  {\Omega
}^{-1}=\Upsilon ( w^{{\prime\prime}},\iota ^{{\prime\prime}}) $
where\ \ $\Upsilon ( w,\iota )  $ are realized by $(2n+2)\times
(2n+2)$ matrices (0). Direct computation then shows that the most
general element with this property is
\begin{equation}
\Omega ( \epsilon ,a,A,w,r) =\left( \begin{array}{ccc}
 a A & 0 & w \\
 -^{t}w \zeta \mbox{}^{\circ} A & \epsilon  a^{2} & r \\
 0 & 0 & \epsilon 
\end{array}\right)  ,%
\label{matrix reps of AutH}
\end{equation}

\noindent where $A\in \mathcal{S}p( 2n) $, $w\in \mathbb{R}^{2n}$,
$a,r \in \mathbb{R}$, $\epsilon =\pm 1$\ \ and $\zeta \mbox{}^{\circ}$
is the symplectic metric defined in (0). The group multiplication
and inverse are
\begin{equation}
\begin{array}{c}
 \begin{array}{rl}
 \Omega ( \epsilon ^{{\prime\prime}},a^{{\prime\prime}},A^{{\prime\prime}},w^{{\prime\prime}},r^{{\prime\prime}})
& =\Omega ( \epsilon ,a,A,w,r) \Omega ( \epsilon ^{\prime },a^{\prime
},A^{\prime },w^{\prime },r^{\prime }) 
\end{array} \\
 =\Omega ( \epsilon  \epsilon ^{\prime },a a^{\prime },A A^{\prime
},\epsilon ^{\prime }w+a A w^{\prime },\epsilon ^{\prime }r+\epsilon
a^{2}r^{\prime }{- }^{t}w \zeta \mbox{}^{\circ} A w^{\prime } )
\\
 {\Omega ( \epsilon ,a,A,w,r) }^{-1}=\Omega ( \epsilon ,a^{-1},a^{-1}A^{-1},-\epsilon
a^{-1}A^{-1}w,-a^{-2}r) 
\end{array}
\end{equation}

\noindent It follows directly that the automorphisms are
\begin{equation}
\begin{array}{rl}
 \Upsilon ( w^{{\prime\prime}},r^{{\prime\prime}})  & =\Omega (
\epsilon ^{\prime },a^{\prime },A^{\prime },w^{\prime },r^{\prime
}) \Upsilon ( w,r) {\Omega ( \epsilon ^{\prime },a^{\prime },A^{\prime
},w^{\prime },r^{\prime }) }^{-1} \\
  & =\Upsilon ( \epsilon ^{\prime }a^{\prime }A^{\prime }w,a^{2}
r-{\epsilon ^{\prime } }^{t}w^{\prime }\cdot \zeta \cdot A^{\prime
}\cdot w + a^{\prime } {}^{t}\left( A^{\prime } w\right) \zeta 
w^{\prime }) 
\end{array}
\end{equation}

\noindent The central extension of the automorphism group of the
Weyl-Heisenberg group is
\begin{equation}
\check{\mathcal{A}}{\mathrm{ut}}_{\mathcal{H}}\simeq {\overline{\mathcal{A}ut}}_{\mathcal{H}}=\left(
\mathcal{A}\otimes \mathcal{Z}_{2}\right) \otimes _{s}\overline{\mathcal{H}\mathcal{S}p}(
2n+2) %
\label{Central extension of AutH}
\end{equation}

\noindent where\ \ $\overline{\mathcal{H}\mathcal{S}p}( 2n+2) \simeq
\overline{\mathcal{S}p}( 2n+2) \otimes _{s}\mathcal{H}(n+1)$. Folland
also shows that the automorphism group for the Weyl-Heisenberg algebra
and group are the same\ \ \cite{folland}.\ \ Note that the central
generator $I$ of the algebra of\ \ $\mathcal{H}( n+1) $ is also
a central element of $\check{\mathcal{A}}{\mathrm{ut}}_{\mathcal{H}}$.\ \ 
\subsection{Invariance of Time}\label{Appendix Matrix group}

The subgroup $\mathcal{G}$ of the homogeneous subgroup of ${\mathcal{A}ut}_{\mathcal{H}}$,
$g\in \mathcal{G}\subset \mathcal{A}\otimes _{s}\mathcal{Z}_{2}\otimes
_{s}\mathcal{S}p( 2n+2) $, such that $g T g^{-1}=T$ is established
directly through a matrix realization of the symplectic group and
Weyl-Heisenberg algebra.\ \ First note that a matrix $S\in \mathcal{S}p(
2n+2) $ leaves invariant the symplectic metric\ \ $S \zeta  {{}
}^{t}S=\zeta $.\ \ The matrix for $\zeta $ and the matrix representation
for the Lie algebra generator $T$ are realized\ \ as\ \ $(2n+2)\times
(2n+2)$ matrices as follows in the basis ordering $\{Z_{\alpha }\}=\{P_{i},Q_{i},E,T\}$\ \ 
\begin{equation}
\zeta =\left( \begin{array}{ccc}
 \zeta \mbox{}^{\circ} & 0 & 0 \\
 0 & 0 & -1 \\
 0 & 1 & 0
\end{array}\right) ,\ \ T=\left( \begin{array}{ccc}
 0 & 0 & 0 \\
 0 & 0 & 1 \\
 0 & 0 & 0
\end{array}\right) .%
\label{zeta and T}
\end{equation}

\noindent $\zeta \mbox{}^{\circ}$ is defined in (0).\footnote{The
relative sign of $\zeta \mbox{}^{\circ}$ and the $2\mathrm{x2}$
symplectic submatrix is a convention. An automorphism takes one
form into the other. Note that $-d e \wedge d t+d p \wedge d q$
$= d t\wedge d e+d p \wedge d q$ by the properties of the wedge
product.}\ \ Using ${S }^{-1}=-\zeta  {{} }^{t}S \zeta $ ,\ \ a
general matrix element\ \ $S$ and $S^{-1}$ of\ \ $\mathcal{S}p(
2n+2) $ have the form \cite{Gosson} 
\begin{equation}
S=\left( \begin{array}{ccc}
 B & b_{1} & b_{2} \\
 c_{1} & d_{1,1} & d_{1,2} \\
 c_{2} & d_{2,1} & d_{2,2}
\end{array}\right)  ,\ \ \ \ S^{-1}=\text{}{\left( \begin{array}{ccc}
 B^{-1} & - \zeta \mbox{}^{\circ} {}^{t}c_{2} & \zeta \mbox{}^{\circ}
{}^{t}c_{1} \\
 {}^{t}b_{2}\zeta \mbox{}^{\circ} & d_{2,2} & -d_{1,2} \\
 {}-^{t}b_{1}\zeta \mbox{}^{\circ} & -d_{2,1} & d_{1,1}
\end{array}\right) }^{-1}.%
\label{Symplectic with Time invariant}
\end{equation}

\noindent $B$ is a $2n\times 2n$ matrix, $b_{\alpha },c_{\alpha
}\in \mathbb{R}^{2n}$ and $d_{\alpha ,\beta }\in \mathbb{R}$, $\alpha
,\beta =1,2$. Imposing the condition $T=S T S^{-1}$ through matrix
multiplication results in $b_{1}=0$, $d_{2,1}=0$ and $d_{1,1}=\epsilon
$, $\epsilon =\pm 1$.\ \ Finally, imposing the condition $S S^{-1}=I_{2n+2}$
results in\ \ $c_{2}=0$, $d_{2,2}=\epsilon $ and $c_{1}=-\epsilon
\ \ {}^{t}b_{2}\zeta \mbox{}^{\circ}A$. Therefore, elements of $\mathcal{S}p(
2n+2) $ that leave $T$ invariant have the form $\Gamma \mbox{}^{\circ}\in
\mathcal{H}\mathcal{S}p( n) $ are given above in (0).\ \ Similar
arguments result in the invariance of ${}^{t}Q_{i}Q_{i}$ to reduce
the form further to $\Phi \mbox{}^{\circ}\in \mathcal{H}a( n) $.\label{sufficient}\label{antiunitary}\label{intrinsic}\label{nr}\label{WH}\label{stone}\label{i
in heisenberg}\label{general aut}\label{vfr dim}\label{zz}\label{limit}\label{zeta}

\end{document}